\documentclass[review,12pt,authoryear]{elsarticle}

\usepackage{epstopdf, epsfig}
\usepackage{amsmath,amssymb, amsfonts, amstext}
\usepackage{natbib}
\usepackage{graphicx}
\usepackage{dcolumn}
\usepackage{bm}
\usepackage{color}
\usepackage{tikz, tikz-3dplot,pgfplots}
\usepackage[overload]{empheq}
\usepackage{float}
\usepackage{color}
\usepackage{lineno}

\newcommand{\parantes}[1]{\left(#1\right)}
\newcommand{\vdot}{\mathop{\mbox{\boldmath{$\cdot$}}}}

\journal{}

\begin{document}

\begin{frontmatter}

\title{Numerical simulation of nonlinear long waves interacting with arrays of emergent cylinders}

\author[label1]{Amir Zainali\corref{cor1}}
\ead{mrz@vt.edu}
\cortext[cor1]{Corresponding author}
\author[label1]{Roberto Marivela}
\author[label1]{Robert Weiss}
\author[label2]{Jennifer L. Irish}
\author[label2]{Yongqian Yang}
\address[label1]{Department of Geosciences, Virginia Tech, Blacksburg, VA 24060, USA}
\address[label2]{Department of Civil and Environmental Engineering, Virginia Tech, Blacksburg, VA 24060, USA}

\address{}

\begin{abstract}
We presented numerical simulation of long waves, interacting with arrays of emergent cylinders inside regularly spaced patches, representing discontinues patchy coastal vegetation. We employed the fully nonlinear and weakly dispersive Serre-Green-Naghdi equations (SGN) until the breaking process starts, while we changed the governing equations to nonlinear shallow water equations (NSW) at the vicinity of the breaking-wave peak and during the runup stage. We modeled the cylinders as physical boundaries rather than approximating them as macro-roughness friction. We showed that the cylinders provide protection for the areas behind them. However they might also cause amplification in local water depth in those areas. The presented results are extensively validated against the existing numerical and experimental data. Our results demonstrate the capability and reliability of our model in simulating wave interaction with emergent cylinders. 
\end{abstract}


\end{frontmatter}


\section{Introduction}
\label{intro}
Tsunamis, caused by different events such as earthquakes and landslides, and storms pose significant threat to human life and offshore and coastal infrastructure. The effects of coastal vegetation on long waves have been investigated extensively in the past \citep[e.g.][]{mei2011long, anderson2014wave}. It has been considered that continuous vegetation provides protection for the areas behind them \citep[e.g.][]{irtem2009coastal, tanaka2007coastal}. However, limited studies focused on the propagation and run-up of long waves in the presence of discontinuous vegetation \citep[see][]{irish2014laboratory}. The following question arises: Do discontinuous arrays of cylinders, representing coastal vegetation such as mangroves, coastal forests, and man-made infrastructure, act as barriers or as amplifiers for energy coming ashore during coastal flooding events?

Three dimensional numerical simulation is the superior choice when computational accuracy is concerned. However, three dimensional models are computationally very expensive, and the large scale of real-world problems limits their applications in practice. Furthermore, the vertical component of the flow acceleration is small compared to the horizontal components. Thus, high fidelity depth integrated formulations, such as nonlinear shallow water equations (NSW), can be an attractive alternative for practical problems.   

\begin{figure}
\begin{center}
\includegraphics[scale=0.75]{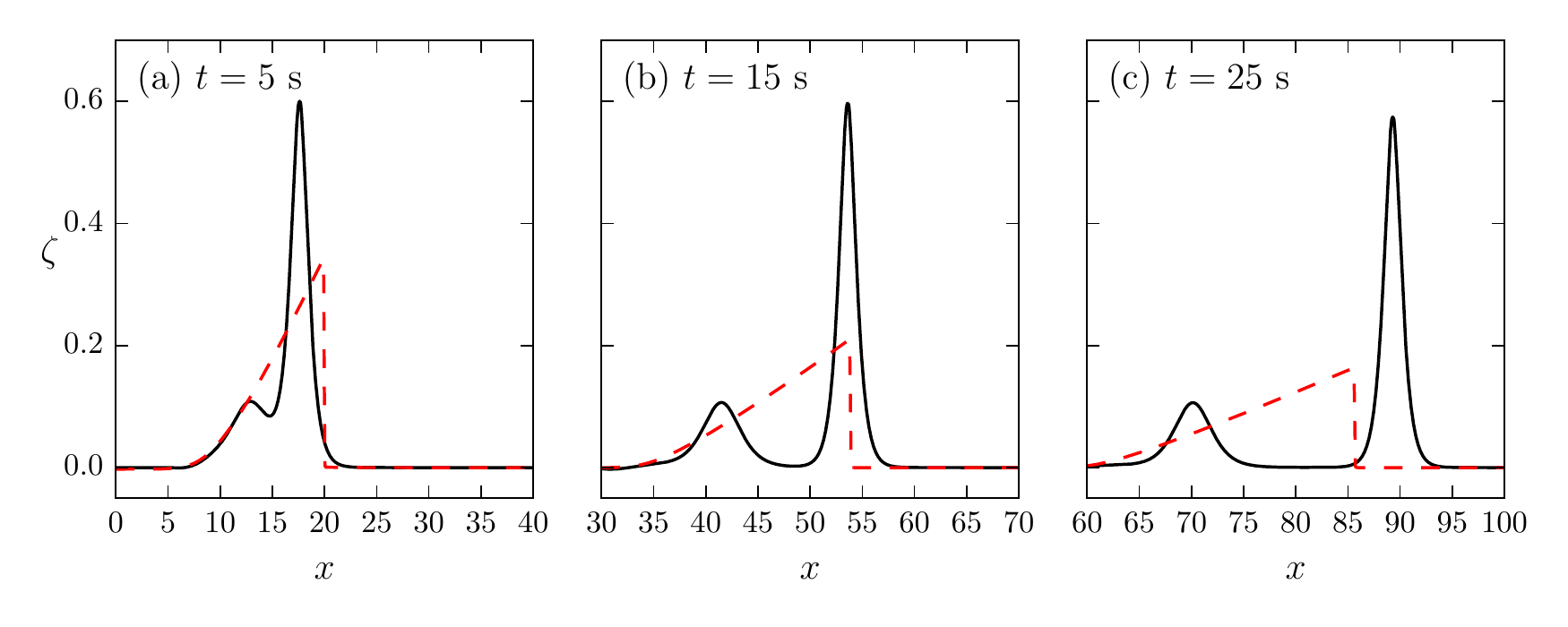}
\end{center}
\caption{\label{fig:sgntonsw} Free surface elevation of the solitary-type transient wave over constant water depth at (a) $t=5$ s, (b) $t=15$ s, and (c) $t=25$ s; (\ref{gnlegend}) 1D-SGN, and (\ref{swlegend}) 1D-NSW. The following parameters are used: $h_0=0.73\;\text{m}$, $H=0.50\;\text{m}$, $k_0 = 0.54\;\text{m}^{-1}$. }
\end{figure}

NSW have been extensively used for simulating long waves. Due to their conservative and shock-capturing properties, they represent a suitable approximation of the wave breaking as well as inundation. However, these equations are only valid for very long waves, and they cannot properly resolve dispersive effects before wave breaking. Thus, they become inaccurate in predicting the effects of shorter wavelength which is important for simulating the nearshore wave characteristics. On the other hand Boussinesq-type equations take dispersive effects into account and can be used to simulate the nearshore wave propagation until the breaking point more accurately. The importance of dispersive effects, until the breaking process starts, is demonstrated in Fig. \ref{fig:sgntonsw}. Using SGN the transient leading wave almost maintains its shape, during propagation over constant water depth. However, if we ignore the effects of dispersive terms (i.e. by using NSW), after traveling a sufficiently large distance (for this scenario 80 m) the wave height decreases by the factor of 3.

Different classes of higher-order depth-integrated equations, based on Boussinesq wave theory, have been derived and presented in literature on simulating shallow water flows: \citet[][]{wei1995fully, liu1994model, lannes2009derivation}.  In this study, we employ fully nonlinear weakly dispersive Boussinesq equations, also known as Serre-Green-Naghdi equations \citep[SGN,][]{lannes2009derivation}. These equations have two advantages compared to others: (a) They only have spatial derivatives up to second order while other equations usually include spatial derivatives of third order, and (b) they do not have any additional source therm in the conservation of mass. These properties improve the computational stability and robustness of the model. Furthermore, unlike earlier studies in which vegetation effects are approximated by an ad-hoc bottom friction coefficient \citep[e.g.][]{yongqiannees, mei2011long}, we model the cylinders as physical boundaries. This enables us to simulate the wave propagation through discontinuous arrays of cylinders without making any assumptions. We demonstrate the capability of the SGN equations in simulating of the waves interacting with both offshore and coastal structures while sustaining the computational performance.

\section{Theoretical background}
\label{theory}

We are interested in simulating long waves; thus $\mu=(h_0/\lambda)^2<<1$, where $\lambda$ denotes the wavelength, and $h_0$ is the still water depth. Assuming an inviscid and irrotationl flow (in the vertical direction), expanding the Euler equations into an asymptotic series and keeping terms up to $O(\mu^2)$, SGN equations can be written in the form of
 
\begin{subequations}
\label{eq:ge}
\begin{align}[left = \empheqlbrace\,]
& \partial_t h + \bm{\nabla}\vdot \parantes{ h \bm{u} } = 0, \label{eq:mass}\\
& \partial_t \parantes{h \bm{u}} + \bm{\nabla}\vdot \parantes{ h \bm{u} \otimes \bm{u} + \frac{1}{2}g h^2 \text{\textbf{I}}  } =  -g h \bm{\nabla} b +  \bm{\mathcal{D}},\label{eq:mom}
\end{align}
\end{subequations}
where $\bm{\mathcal{D}}$ is a nonlinear function of the free-surface elevation $\zeta$, the depth-integrated velocity vector $\bm{u}$, and their spacial derivatives; $h$ is the water depth; $b$ represents the bottom variations; and $\text{\textbf{I}}$ is the identity tensor. See \citet{bonneton2011splitting} and \citet{lannes2009derivation} for the complete form of the equations and their derivation. Note that by setting $\bm{\mathcal{D}}=0$, Eqs. (\ref{eq:mass})-(\ref{eq:mom}) will reduce to the NSW equations (accurate to $O(\mu)$). 

We employ the Basilisk code to solve the governing equations \citep[Basilisk, URL: basilisk.fr,][]{popinet2015quadtree}. The computational domain is discretized into a rectangular grid and solved using the second order accurate scheme in time and space. We refer to \citet{popinet2015quadtree} for more information about the numerical scheme. Friction effects become important as the wave approaches the coast. Following \citet{bonneton2011splitting}, we added a quadratic friction term $f=c_f\frac{1}{h}|\bm{u}|\bm{u}$ to Eq. (\ref{eq:mom}) where the friction coefficient is $c_f=0.0034$. 

\begin{figure}
\begin{center}
\includegraphics[scale=1.1]{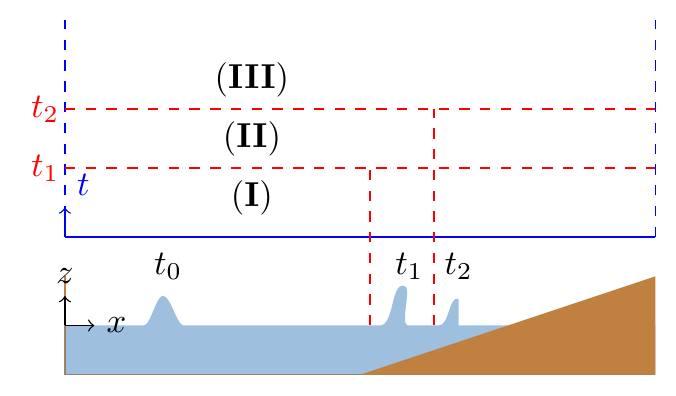}
\end{center}
\caption{\label{fig:NEES} Schematic of the breaking process. The vertical dashed lines indicate the boundary of subdomains. The governing equations in the left subdomain are SGN and NSW equations elsewhere. The boundary follows the leading wave; (\textbf{I}) $t<t_1:\;\;$SGN in the whole domain, (\textbf{II}) $t_1<t<t_2:\;\;$SGN in the left subdomain and SW in the right subdomain, and (\textbf{III}) $t>t_2:\;\;$NSW in the whole domain.}
\end{figure}  

\subsection{Initial and boundary conditions}

We imposed the initial conditions
\begin{equation}
\label{bczeta}
\zeta\parantes{x}|_{t=0} = H \textrm{ sech}^2 \left[ k_0 (x-x_0)\right], \; \bm{u}(x)|_{t=0} = \frac{c\zeta}{\zeta+ h_0}, 
\end{equation}
to generate solitary-type transient long waves \citep{madsen2008solitary, madsen2010analytical}, in which $c = \sqrt{g(h_0+H)}$ is the phase speed.  $H$ represents the initial wave height, $c$ denotes the phase speed, and $k_0$ is the characteristic wave number. This wave type removes the constraint that exists between the wave height and the wave length and will result in a more realistic representation of tsunami waves \citep[e.g.][]{baldock2009kinematics, rueben2011optical, irish2014laboratory} By choosing $k_0 = \sqrt{3H/4h^3_0}$, Eq. (\ref{bczeta}) will reduce to the free-surface elevation of a solitary wave with a permanent shape.

\subsection{Wave breaking}
\label{sec:breaking}
We utilize the conditions suggested in \cite{tissier2012new} to estimate the start time of the breaking process. As noted earlier, NSW are employed as governing equations in the vicinity of the breaking-wave peak while modeling the rest of the domain using SGN. Here, we divide the computational domain into two sub-domains. The boundary that separates these two sub-domains follows the leading breaking wave (see Fig. \ref{fig:NEES}). We solve the domain behind the wave peak using SGN and the rest of the domain using NSW ($t_1<t<t_2$ in Fig. \ref{fig:NEES}). The distance between the wave peak and the boundary is given by $x_b-x_w = 2h|_{x_w}$ where $x_b$ is the coordinate of the boundary and $x_w$ is the coordinate of the wave peak.

Note that in the very shallow areas, i.e., in the vicinity of the shoreline and the coastal areas, we always model the flow using NSW equations to avoid the numerical dispersions that can be caused by dispersive terms in SGN. In this study, we considered the run-up of a single wave. After the wave breaks completely ($t>t_2$ in Fig. \ref{fig:NEES}), to increase the computational efficiency, we model the entire domain using NSW. 

\section{Results}
\subsection{Validation of numerical results}
\subsubsection{Non-breaking solitary wave interaction with a group of cylinders}
Solitary wave interaction with a group of three vertical cylinders is investigated by \cite{mo2009three} in which they performed three-dimensional numerical simulations using volume of fluid method (VOF) and compared their numerical results to experiments. We compare our results from two-dimensional SGN with the experiments and the three-dimensional VOF simulations by \citet{mo2009three}. A solitary wave with $H=0.3h_0$, $k_0 = \sqrt{3H/4h^3_0}$, and $x_0=15\;\text{m}$ inside a rectangular domain of $(48.8 \;\text{m}, 26.5 \;\text{m})$ interacts with three cylinders. The diameter of cylinders is $1.22 \;\text{m}$ and they  are located at $(27.7 \;\text{m}, 0 \;\text{m})$, $(30.1 \;\text{m}, 1.2 \;\text{m})$ and $(30.1\;\text{m}, -1.2 \;\text{m})$. The still water depth is $h_0=0.75 \;\text{m}$. Uniform rectangular grid with $\Delta x/L_0=1024 $ is used where $L_0=48.8 \;\text{m}$ is the length of the computational domain. 

\begin{figure}
\begin{center}
\includegraphics[scale=0.9]{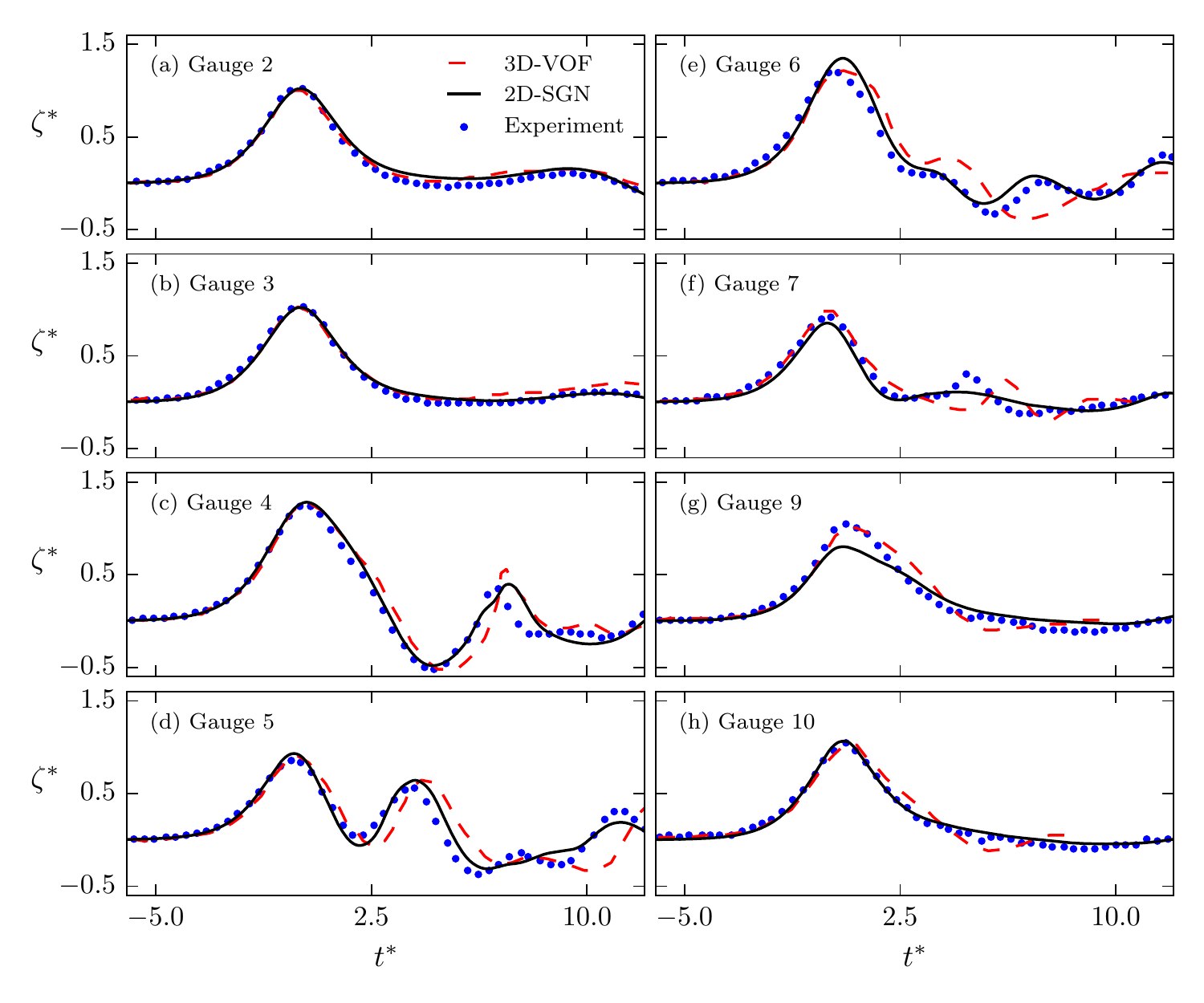}
\end{center}
\caption{\label{fig:mo} Comparison of numerical and experimental data presented. Experimental and three-dimensional simulations are from \cite{mo2010numerical}; $t^*=t/\sqrt{h/g}$, and $\zeta^*=\zeta/H$; (\ref{gnlegend}) present simulation (2D-SGN), (\ref{explegend}) experimental results and (\ref{numlegend}) three-dimensional simulation (3D-VOF). }
\end{figure}

Figure \ref{fig:mo} depicts the comparison of the wave elevation at different wave gauges obtained using SGN with those of \cite{mo2010numerical}. The results are in excellent agreement except for the wave gauges immediately in front and back of the first cylinder (see Fig. \ref{fig:mo}(e) and \ref{fig:mo}(g)). Our model overestimates the leading wave height in front and underpredicts the wave height in back of the first cylinder. Similar errors are reported between experiments and Boussinesq models in \cite{zhong2009modeling}. These differences are due to the fact that the three-dimensional effects of the flow become more relevant in close proximity to the cylinders. Using SGN, secondary wave elevations are closer to the experimental data than the three dimensional simulation. This demonstrates the capability of SGN for simulating the effects of shorter waves that can become very important in the near shore regions.

\subsubsection{Breaking solitary-type transient wave run-up}
\label{sec:runup}

We investigated solitary-type transient wave run-up on a steep sloping beach (1:10) following the experimental setup in \citet{irish2014laboratory}. We discretized the computational domain (52 \;\text{m}, 4.4 \;\text{m}) using a rectangular uniform grid with $\Delta x = \Delta y =52/1024\;\text{m}$. The toe of the sloping beach is located at $x = 32 \;\text{m}$. The still water depth is $h_0=0.73 \;\text{m}$. The transient solitary wave is generated by setting $H=0.50\;\text{m}$, $k_0 = 0.54\;\text{m}^{-1}$, and $x_0=10 \;\text{m}$ at $t=0\;\text{s}$. A symmetry boundary condition is imposed on bottom and top boundaries.

\begin{figure}
\begin{center}
\includegraphics[scale=0.9]{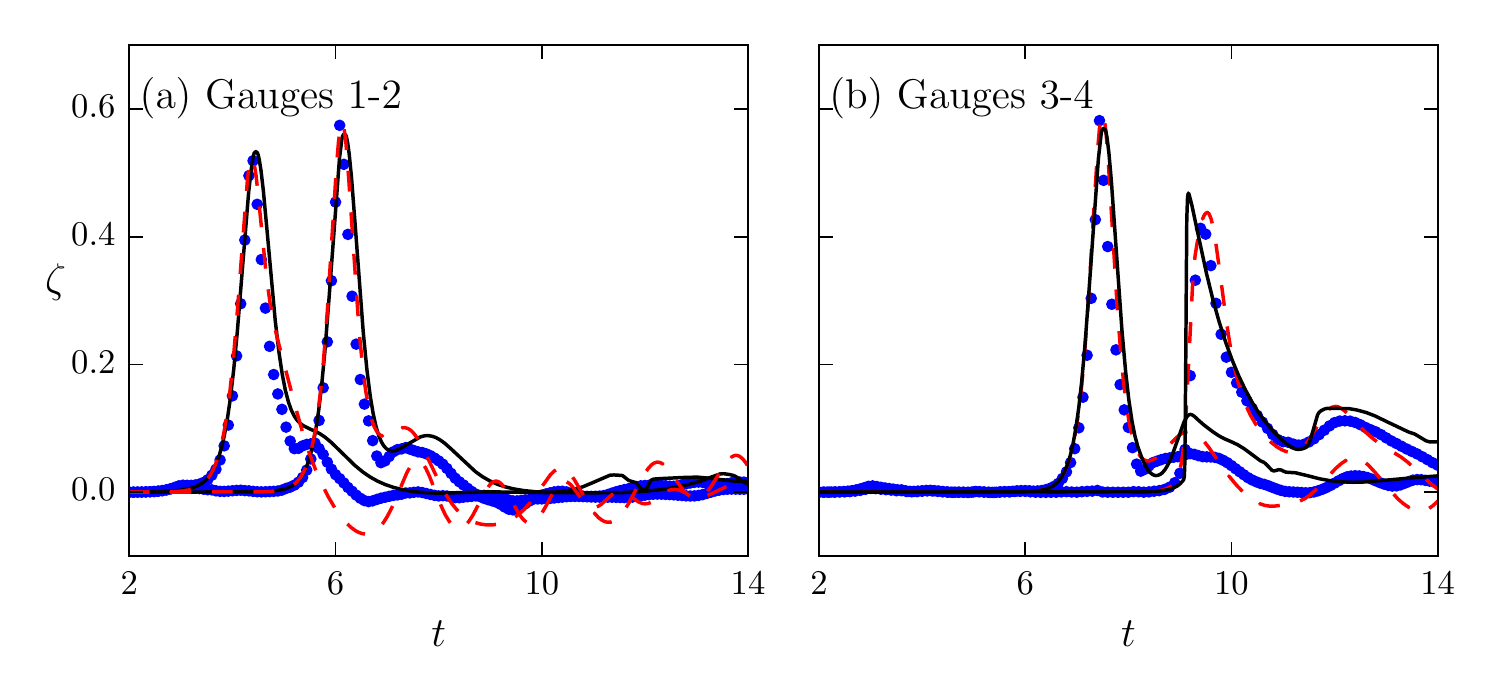}
\end{center}
\caption{\label{fig:offshore} Free-surface elevation at (a) gauges 1-2, and (b) gauges 3-4. Experimental and COULWAVE results are summarized in \cite{yongqiannees}; (\ref{gnlegend}) present simulation (2D-SGN), (\ref{explegend}) experimental results and (\ref{numlegend}) COULWAVE. }
\end{figure}

\cite{yongqiannees} studied this problem experimentally as well as numerically using COULWAVE \cite[Cornell University Long and Intermediate Wave Modeling Package,][]{lynett2002modeling} which solves the Boussinesq equations described in \cite{liu1994model} and \cite{lynett2002modeling}. We compared the free surface elevation (wave gauges 1-4) and the local water depth (wave gauges 5-16) with the results presented in \cite{yongqiannees}. The wave gauge coordinates are summarized in Tab. \ref{tab:1}. Figure \ref{fig:offshore} shows the free-surface elevation $\zeta$ at gauges 1-4. While both methods capture the evolution of the free surface accurately, we observe that COULWAVE captures the breaking process slightly better. This can be due to the differences in the how wave breaking is handled. COULWAVE employs an ad-hoc viscosity model in the breaking process. In contrast, we solve NSW in the breaking zone and incorporate a shock-capturing scheme in the breaking process. Consequently, our model develops a steeper wave profile.

%

\begin{table}

  \begin{center}
  \begin{tabular}{llclclc}
       \hline
              &  & Location (m) & &  Location (m) & &  Location (m)  \\ [3pt]
        Gauge & 1:&   (18.24, 0.00) & 2:&   (24.43, 0.00) & 3:&   (29.28,  0.00)   \\ 
        Gauge & 4:&   (36.42, 0.00) & 5:&   (46.75, 0.55) & 6:&   (45.10,  2.18)   \\
        Gauge & 7:&   (44.34, 0.83) & 8:&   (48.40, 2.18) & 9:&   (45.65,  0.60)   \\  
        Gauge & 10:&  (49.52, 1.10) & 11:&  (47.30,  2.18)  & 12:&  (47.30, 0.00)  \\
        Gauge & 13:&  (42.90, 0.00) & 14:&  (45.10, 0.00) & 15:&  (47.85,  1.65)   \\
        Gauge & 16:&  (46.20, 1.10) &    &                &    &                  \\ \hline              
  \end{tabular}
  \end{center}
\caption{Wave gauge coordinates.}
\label{tab:1}
\end{table}

\begin{figure}
\begin{center}
\includegraphics[scale=0.70]{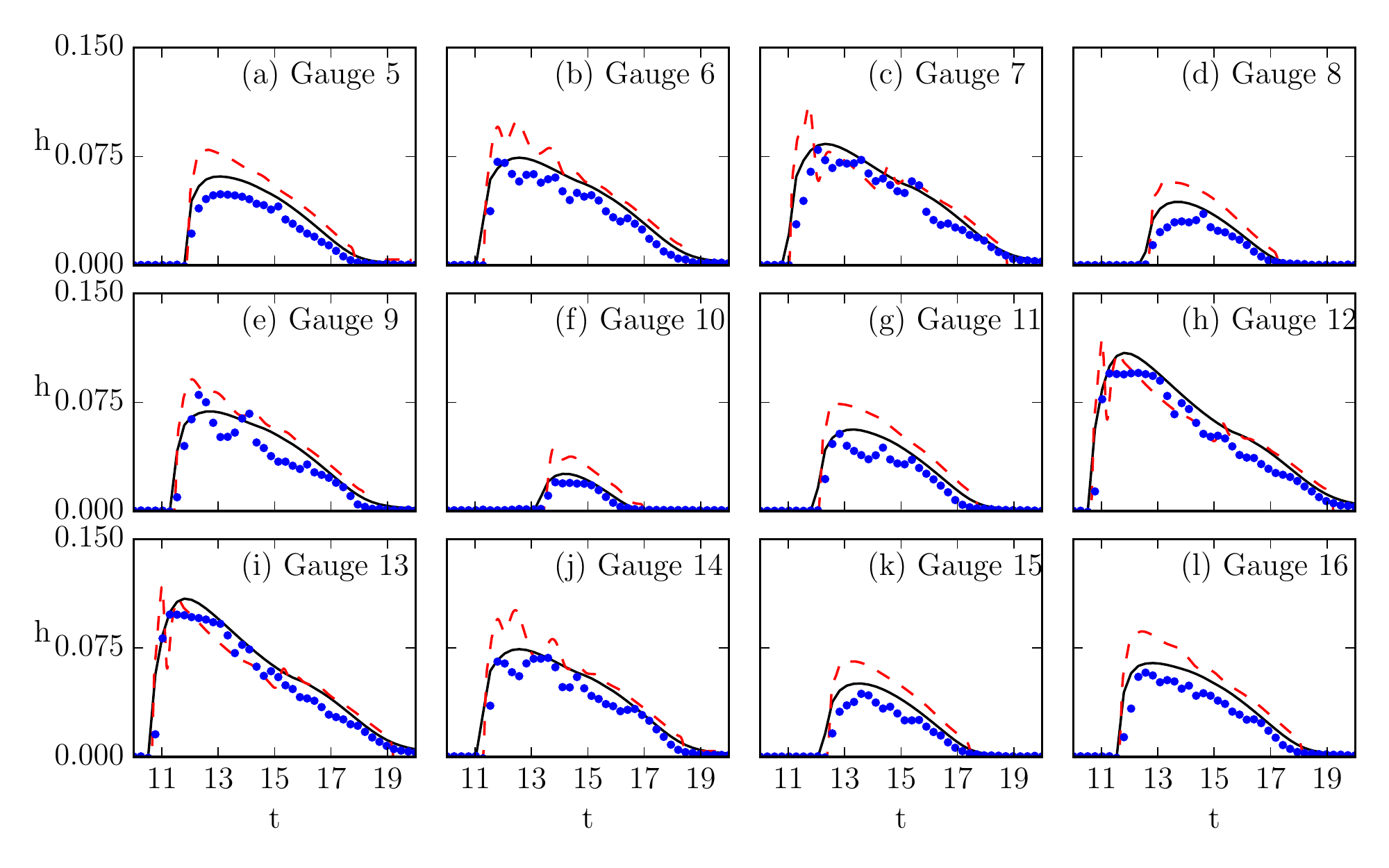}
\end{center}
\caption{\label{fig:allsonic_c} Local water depth at gauges 5-16. Experimental and COULWAVE results are summarized in \cite{yongqiannees}; (\ref{gnlegend}) present simulation (2D-SGN), (\ref{explegend}) experimental results and (\ref{numlegend}) COULWAVE. }
\end{figure}

Figure \ref{fig:allsonic_c} depicts the local water depth at onshore gauges 5-16. Our results are systematically closer to the experimental data. \cite[][]{park2013tsunami} also reported about overpredicting the experimental results using COULWAVE. \cite{yongqiannees} argued that the difference between the simulation and the experiment can be due to the existence of an slight leakage from the sloping beach among some other possible reasons. This can explain the slight overpredictions we observed using our model. 

\begin{table}

  \begin{center}
  \begin{tabular}{lcccccc}
       \hline
         & $d_{r}\;\text{(m)}$ & $N_{cp}$ & $d_p\;\text{(m)}$ &  $C_{fp}\;\text{(m)}$ & $D_c\;\text{(m)}$ &$D_p\;\text{(m)}$ \\ [3pt] 
         Scenario 1 &  0.1885 & 21 & 2.2 & (34, 0) & 0.01333 & 0.6 \\  
         Scenario 2 &  0.0943 & 69 & 2.2 & (34, 0) & 0.01333 & 0.6 \\
         Scenario 3 &  0.1885 & 129 & 2.2 & (34, 0) & 0.01333 & 0.6 \\ 
       \hline       
  \end{tabular}
  \end{center}
\label{tab:2}
\caption{Geometrical parameters of macro-roughness patches. $d_r:$ distance between two horizontally (or vertically)  aligned cylinders inside a patch; $N_{cp}:$ total number of cylinders inside a patch; $d_p:$ distance between two horizontally (or vertically)  aligned patches; $C_{fp}:$ coordinate of the center of the first patch; $D_c:$ diameter of the cylinders; $D_p:$ diameter of the patches.}
\end{table}  

\subsection{Breaking solitary-type transient wave run-up in the presence of macro-roughness}

\begin{figure}
\begin{center}
\includegraphics[scale=0.9]{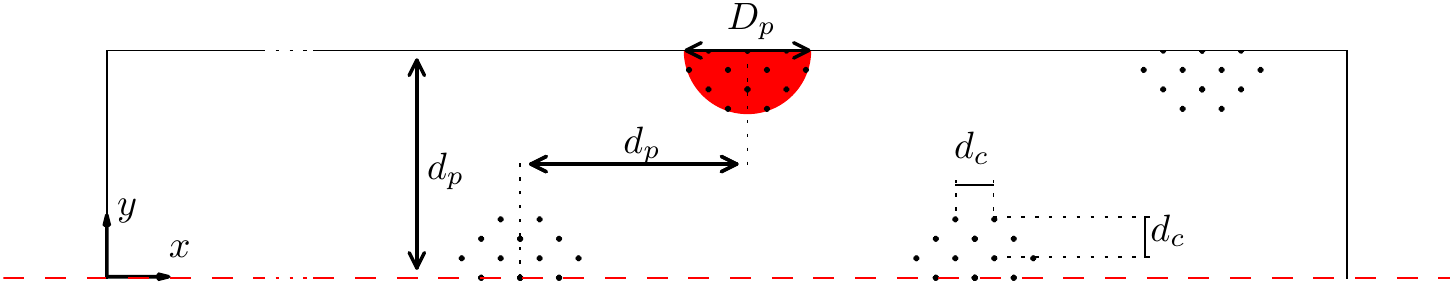}
\end{center}
\caption{\label{fig:patch} Sketch of the macro-roughness patches. }
\end{figure}

A schematic sketch of the problem is shown in Fig. \ref{fig:patch}. The computational domain, boundary conditions, and the initial conditions are the same as the ones in section \ref{sec:runup}. Three different scenarios are considered here. Each macro-roughness patch consists of regularly spaced vertical cylinders. The geometrical parameters of the patches for each scenario are summarized in Tab. \ref{tab:2}. We used a nested mesh with

\begin{align*}
&\Delta x = \Delta y = 52/1024\;\text{m} \;\;\;\;\; x<41\;\text{m}, \\
&\Delta x = \Delta y = 52/8192\;\text{m} \;\;\;\;\; x>41\;\text{m}.
\end{align*}

\begin{figure}
\begin{center}
\includegraphics[scale=0.7]{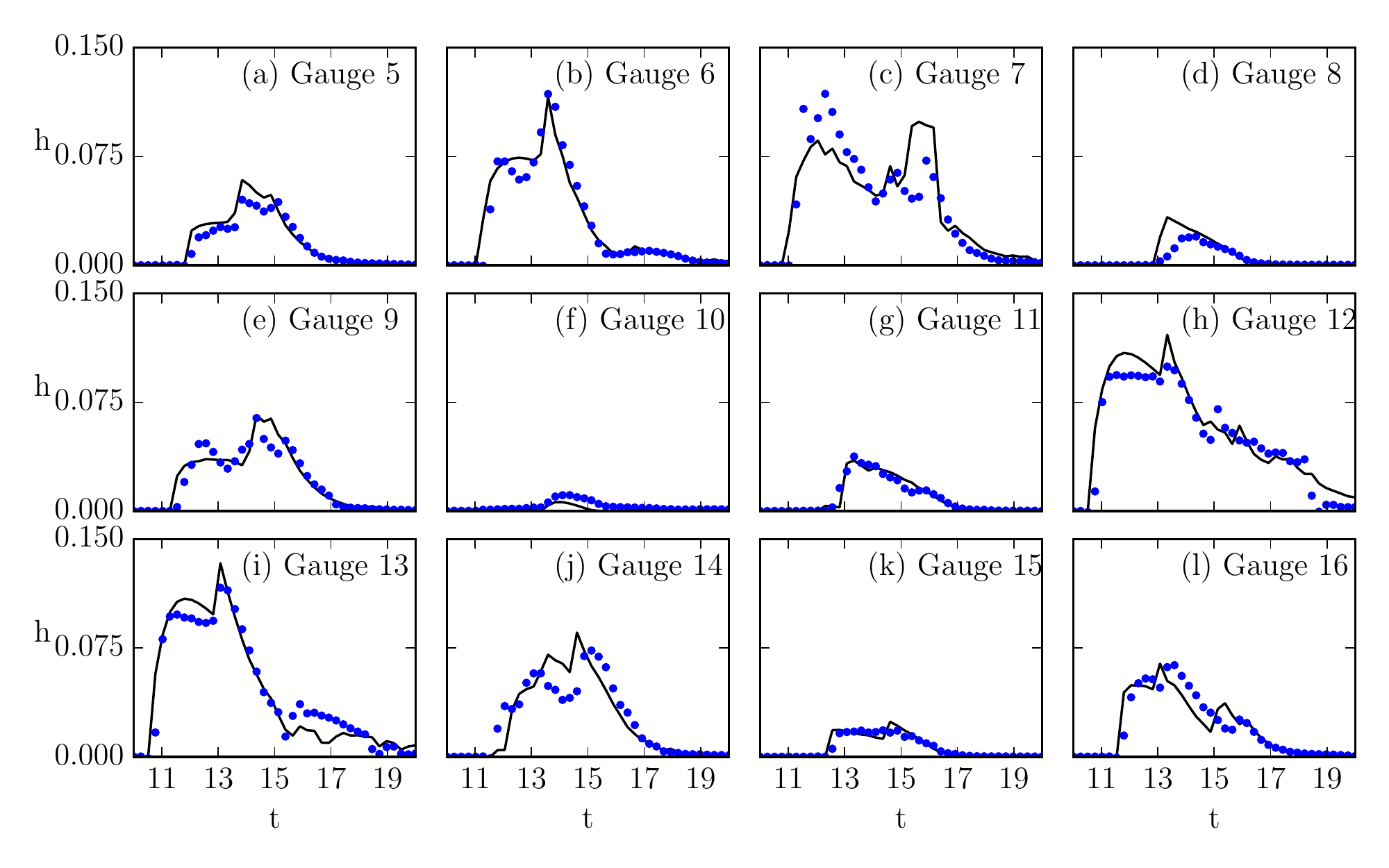}
\end{center}
\caption{\label{fig:allsonic_s6} Local water depth at gauges 5-16 for Scenario 3. Experimental results are summarized in \cite{irish2014laboratory}; (\ref{gnlegend}) present simulation (2D-SGN), and (\ref{explegend}) experimental results. }
\end{figure}  

Local water depth at gauges 5-16 is shown in Fig. \ref{fig:allsonic_s6}. The simulations are in good agreement with the experimental data. Some differences in transient wave peaks can be due to the ensemble-averaged experimental data. Ensemble averaging can smooth out some of the sharp transitions as suggested by \citet{yongqiannees} and \citet{baldock2009kinematics} 

\subsection{Effects of macro-roughness on the local maximum local water depth}

\begin{figure}
\begin{center}
\includegraphics[scale=0.7]{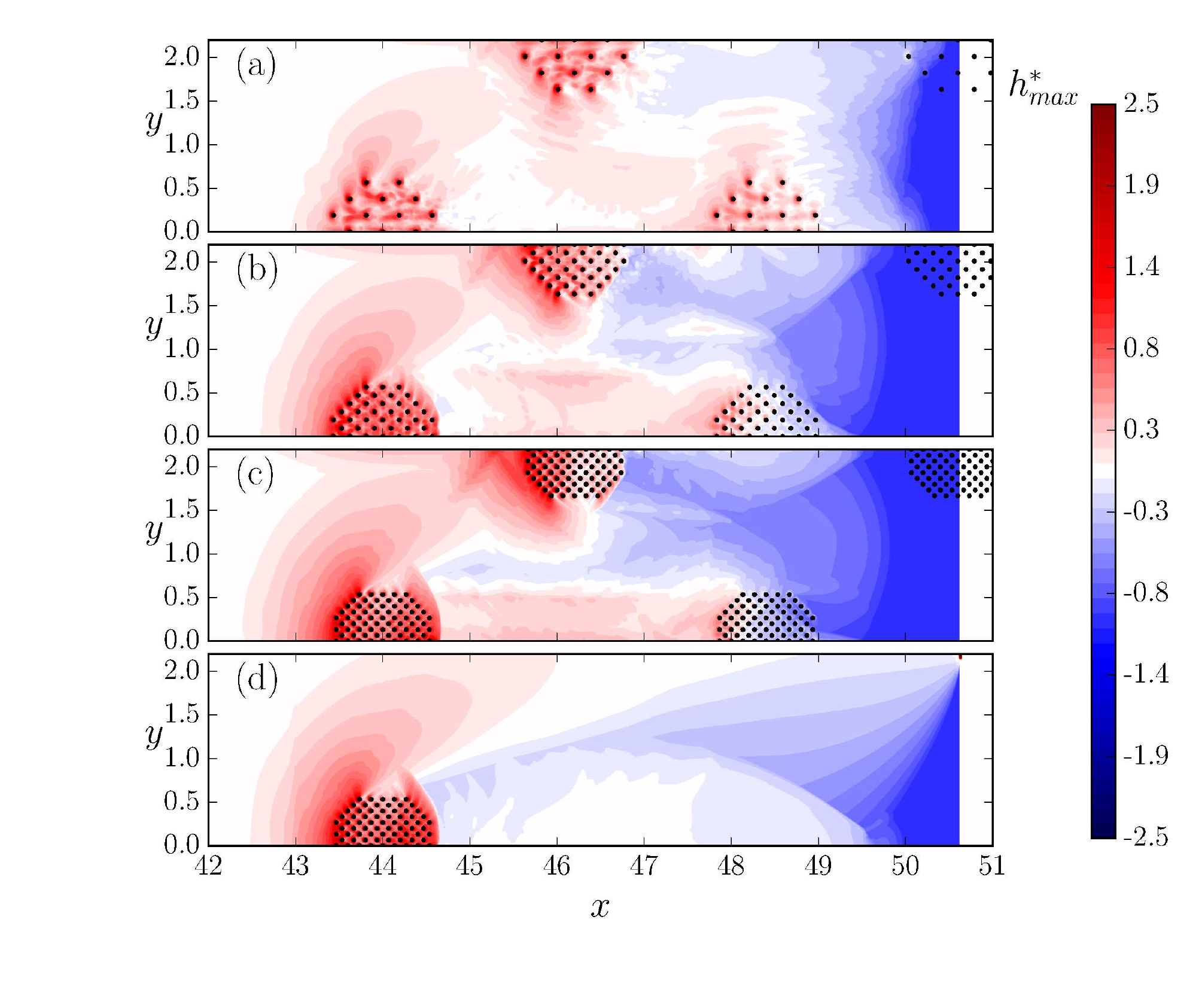}
\end{center}
\caption{\label{fig:hmax} Maximum local water depth $h^*_{max}$ for (a) Scenario 1, (b) Scenario 2 (c) Scenario 3, and (d) Scenario 3 in which all the patches are removed except the first patch. The maximum water depth for each scenario is normalized with the reference values in the absence of macro-roughness patches.  }
\end{figure}

The maximum local water depth is given by

\begin{equation}
h_{max} = \frac{\max(h)-\max(h)|_{ref}}{\max(h)|_{ref}},
\end{equation} 
where $\max(h)|_{ref}$ is the maximum water depth in the absence of the macro-roughness patches. Unlike momentum flux we do observe water depth amplification up to 1.7 times behind the first patch in the presence of the macro-roughness patches. When the flow reaches the first patch the flow refracts away from the center of the patch toward the other patches. However, the patches located in the second row refract and reflect the water toward the centerline. This process causes the water to amplify behind the first patch (red area in Fig. \ref{fig:hmax}). As we can see in Fig. \ref{fig:hmax}(d), no amplification in local water depth occurs behind the patch in the case were all other patches are removed. 

\subsection{Effects of macro-roughness on the local maximum momentum flux}

\begin{figure}
\begin{center}
\includegraphics[scale=0.7]{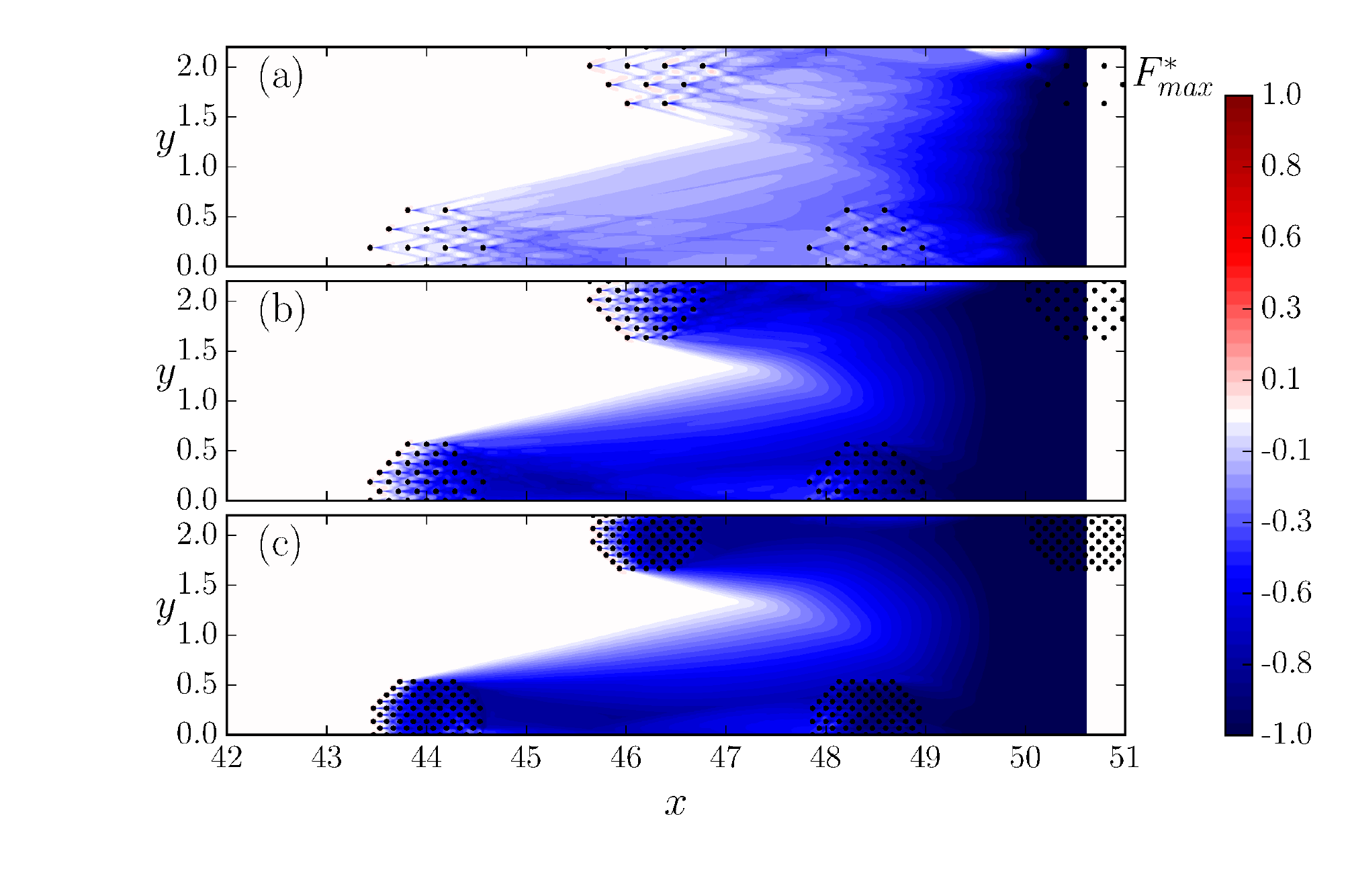}
\end{center}
\caption{\label{fig:mom} Maximum momentum flux $F^*_{max}$ for (a) Scenario 1, (b) Scenario 2, and (c) Scenario 3. The momentum flux for each scenario is normalized with the reference values in the absence of macro-roughness patches. }
\end{figure}

The momentum flux represents the destructive forces of the incident wave. To study the effects of the macro-roughness patches on momentum flux we defined the maximum normalized momentum flux as

\begin{equation}
F_{max} = \frac{\max(h|\bm{u}|^2)-\max(h|\bm{u}|^2)|_{ref}}{\max(h|\bm{u}|^2)|_{ref}},
\end{equation} 
where $\max(h|\bm{u}|^2)|_{ref}$ is the maximum momentum flux in the absence of the macro-roughness patches. Figure \ref{fig:mom} shows the $F_{max}$ for Scenarios 1-3. We can see that the patches provide protection for the areas behind them for Scenario 1 (Fig. \ref{fig:mom}(a)). With increasing density of the cylinders in patches, $F_{max}$ is decreased. Thus the level of protection against incident waves increases. However we did not observe any significant changes with further increasing the density of the cylinders (see Fig. \ref{fig:mom}).

\ 

\subsection{Maximum run-up}

\begin{figure}
\begin{center}
\includegraphics[scale=0.7]{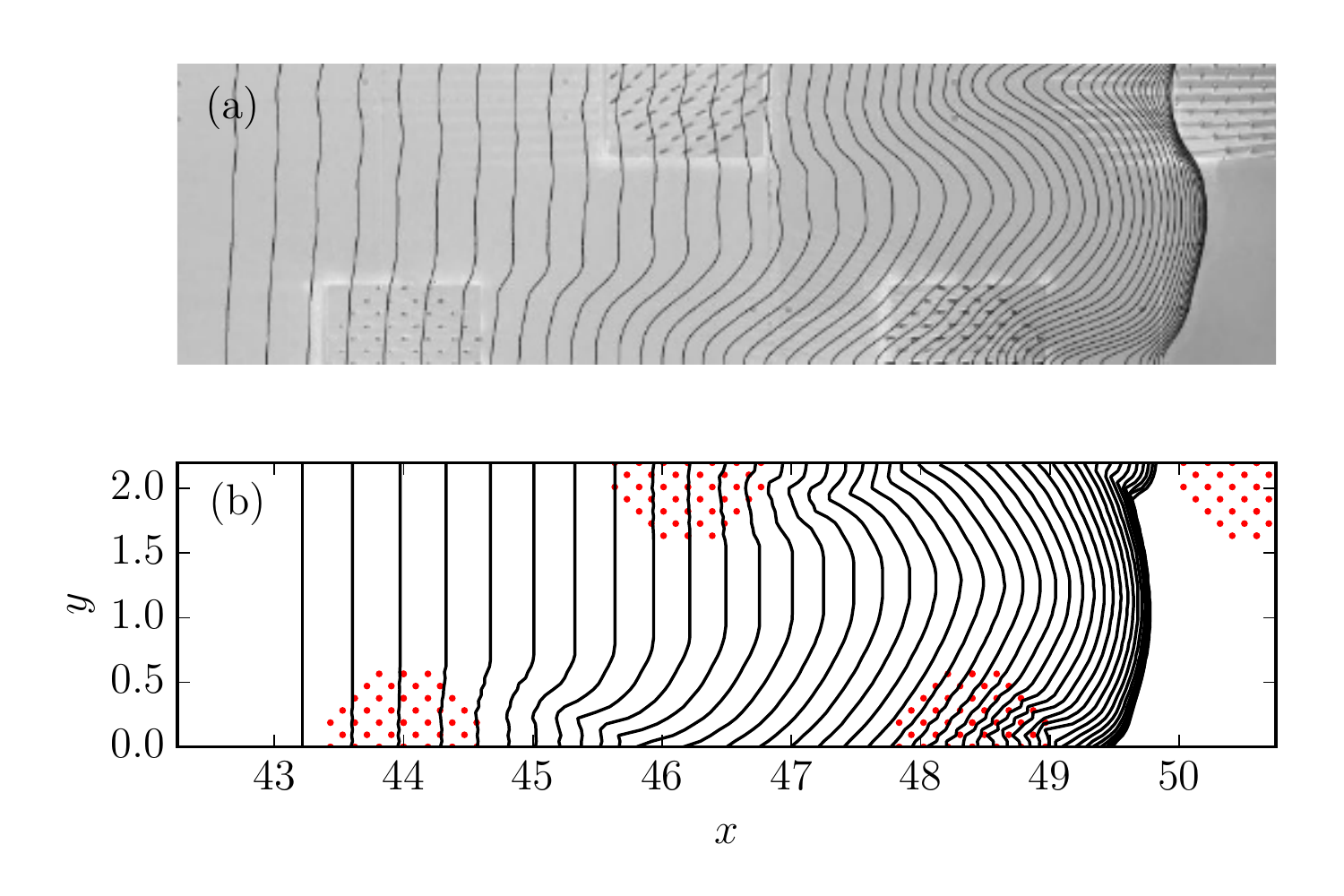}
\end{center}
\caption{\label{fig:bore} Propagation of bore-lines in the presence of macro-roughness patches (Scenario 2).   \cite{irish2014laboratory}; (a) experimental results, and (b) present simulation (2D-SGN).  }
\end{figure}

Maximum run-up is another important criteria in determining the effectiveness of the macro-roughness patches in mitigating tsunami hazard risks. Figure \ref{fig:bore} shows the comparison of the simulation to the experimental results of the bore line propagation for the Scenario 2. The results are in good agreement. The maximum deviation from the experimental results is less than $4\%$.

\begin{figure}
\begin{center}
\includegraphics[scale=0.7]{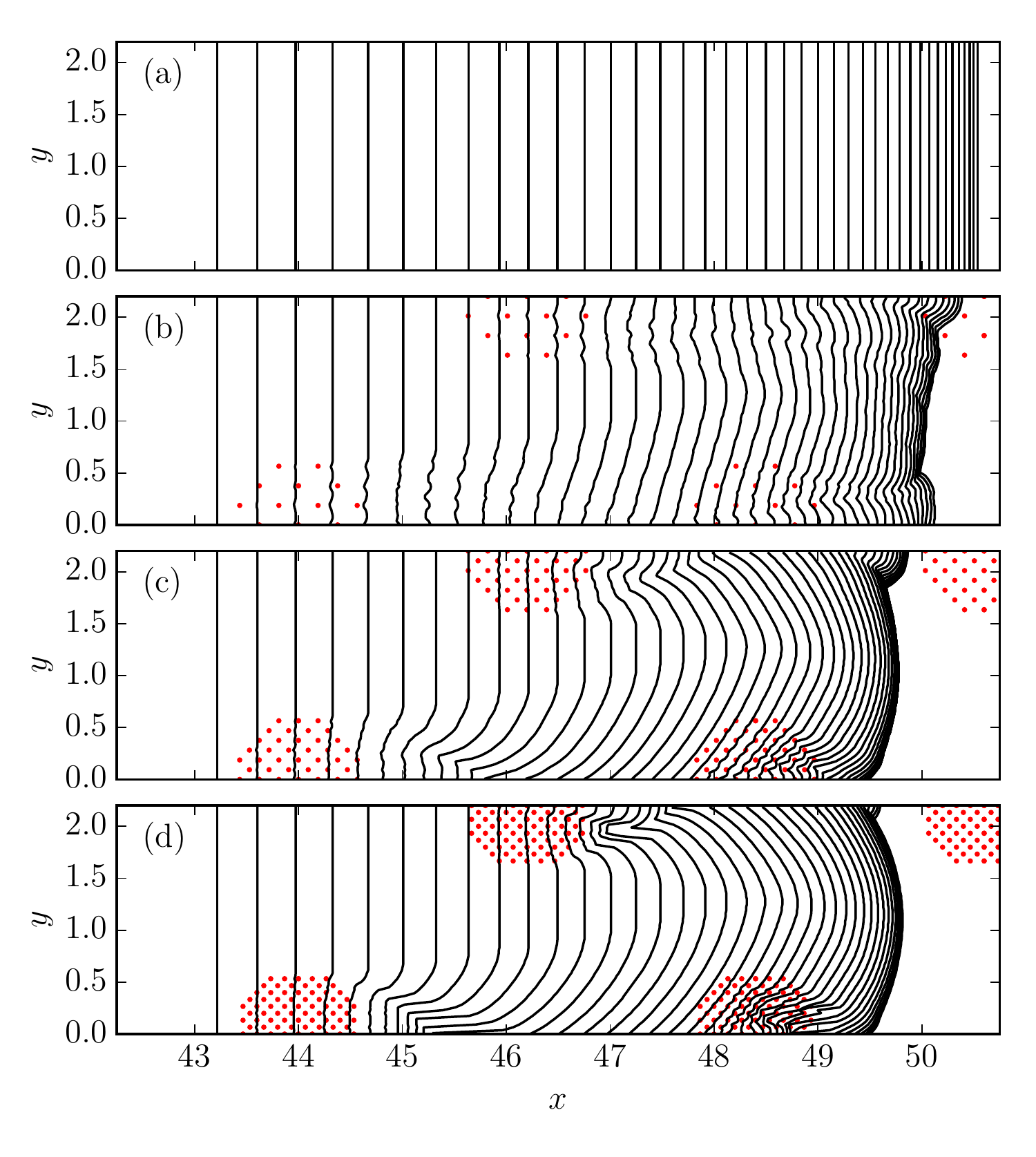}
\end{center}
\caption{\label{fig:bore2} Propagation of bore-lines for (a) Scenario with no macro-roughness patches (b) Scenario 1, (c) Scenario 2 (d) Scenario 3. }
\end{figure}

Fig. \ref{fig:bore2} shows the bore-line propagation for different Scenarios. The maximum run-up decreases with increasing the vegetation density inside the macro-roughness patches. For the very low density vegetations (Fig. \ref{fig:bore2}(b)) this reduction is more or less uniform along the shore. However with increasing the vegetation density (Fig. \ref{fig:bore2}(c)) we observe more reduction behind the patches. With further increasing the vegetation density (Fig.\ref{fig:bore2}(d) in comparison to Fig.\ref{fig:bore2}(c)) while the maximum run-up behind the patches continues to decrease, it increases within the channel between the patches slightly. The reason for this increase is the level by which the flow is channelized between the patches. 

\section{Discussion and conclusion}

We presented numerical simulations of long water waves interacting with emergent cylinders. We demonstrated that higher order depth integrated equations, such as SGN, are a suitable tool to simulate the wave interaction with emergent cylinders accurately except in very close vicinity of the cylinders. Three-dimensional effects cannot be ignored at the proximity of the cylinders and we argue that they are the main reason for the existing differences between results from our model and the experimental data.

Cylinders, representing coastal vegetation, are usually approximated as macro-roughness friction \citep[e.g.][]{yongqiannees, mei2011long}. However there is no analytical solution for a correlation between macro-roughness patterns and the implemented friction coefficient. Thus these models need to be calibrated against available experimental data. However for scenarios with no experimental prototype these models can become inaccurate. Here, with further refining the grid in the coastal areas, we modeled these macro-roughness patches as physical boundaries. 

We observed that discontinuous coastal vegetation can provide protection for the areas behind them. Friction forces become dominant when the flow becomes shallower. In addition, for the areas located on the onshore slope we also have the gravitational force acting in the negative flow direction. Macro-roughness patches elongate the path of the incident wave, causing a longer local inundation period. Thus the flow will be subjected to the negative gravitational and friction force for a longer time and the maximum recorded velocity will decline in comparison to the scenario where no macro-roughness patches exist. This will lead to a reduced local momentum flux and increased protection against the destructive wave force. However, in terms of local maximum water depth this conclusion does not apply. Even though the decreased local velocity will result in more protection against the wave, we observed for the studied macro-roughness the maximum local water depth actually increases behind the patches. We observed amplifications up to 1.6 times in local water depth. This will increase the chance of that area being flooded. We note here that the main reason the Fukushima Daiichi nuclear disaster happened was the fact that the water overtopped the protecting wall and reached the electric generators \citep{synolakis2015fukushima}.     

We demonstrated the capability of the model in analyzing the wave interaction with coastal structures. However, our study was limited to the emergent cylinders and simple macro-roughness patch patterns. Studying submerged coastal vegetations and different patch patterns will be the direction of our future studies.

\section*{References}
\bibliographystyle{elsarticle-harv} 
\bibliography{references.bib}


\scalebox{0.01}{%
\begin{tikzpicture}
    \begin{axis}[hide axis]
        \addplot [only marks,
        color=blue,
        mark options={solid},
        forget plot
        ]
        (0,0);\label{explegend}    
        \addplot [
        color=red,
        dashed,
        line width=1.2pt,
        forget plot
        ]
        (0,0);\label{numlegend}
        \addplot [
        color=black,
        solid,
        line width=1.2pt,
        forget plot
        ]
        (0,0);\label{gnlegend}
       \addplot [
        color=red,
        dashed,
        line width=1.2pt,
        forget plot
        ]
        (0,0);\label{swlegend}        
                
    \end{axis}
\end{tikzpicture}%
}

\end{document}